\documentclass[10pt, conference, letterpaper]{IEEEtran}
\IEEEoverridecommandlockouts
\usepackage{cite}
\usepackage{amsmath,amssymb,amsfonts,amsthm}
\usepackage{algorithmic}
\usepackage[ruled]{algorithm2e}
\usepackage{graphicx}
\usepackage{textcomp}
\usepackage{xcolor}
\usepackage[caption=false,font=footnotesize]{subfi
g}
\def\BibTeX{{\rm B\kern-.05em{\sc i\kern-.025em b}\kern-.08em
    T\kern-.1667em\lower.7ex\hbox{E}\kern-.125emX}}
\usepackage{amsmath}
\DeclareMathOperator*{\argmax}{arg\,max}

\newtheorem{proposition}{Proposition}
\newtheorem{corollary}{Corollary}
\usepackage{xcolor}

\begin{document}

\title{Intelligent Communication Planning \\ for Constrained Environmental IoT Sensing \\ with Reinforcement Learning}
\author{}
\author{\IEEEauthorblockN{Yi Hu}
\IEEEauthorblockA{
\textit{Carnegie Mellon University}\\
yihu@andrew.cmu.edu}
\and
\IEEEauthorblockN{Jinhang Zuo}
\IEEEauthorblockA{
\textit{Caltech/UMass-Amherst}\\
jhzuo@cs.umass.edu}
\and
\IEEEauthorblockN{Bob Iannucci}
\IEEEauthorblockA{
\textit{Carnegie Mellon University}\\
bob@rail.com}
\and
\IEEEauthorblockN{Carlee Joe-Wong}
\IEEEauthorblockA{
\textit{Carnegie Mellon University}\\
cjoewong@andrew.cmu.edu}
}

\maketitle

\begin{abstract}
Internet of Things (IoT) technologies have enabled numerous data-driven mobile applications and have the potential to significantly improve environmental monitoring and hazard warnings through the deployment of a network of IoT sensors.
However, these IoT devices are often power-constrained and utilize wireless communication schemes with limited bandwidth. Such power constraints limit the amount of information each device can share across the network, while bandwidth limitations hinder sensors' coordination of their transmissions. 
In this work, we formulate the communication planning problem of IoT sensors that track the state of the environment. We seek to optimize sensors' decisions in collecting environmental data under stringent resource constraints. 
We propose a multi-agent reinforcement learning (MARL) method to find the optimal communication policies for each sensor that maximize the tracking accuracy subject to the power and bandwidth limitations. MARL learns and exploits the spatial-temporal correlation of the environmental data at each sensor's location to reduce the redundant reports from the sensors.
Experiments on wildfire spread with LoRA wireless network simulators show that our MARL method can learn to balance the need to collect enough data to predict wildfire spread with unknown bandwidth limitations. 
\end{abstract}


\section{Introduction}\label{sec:intro}

\begin{figure}[t]
    \centering
    \includegraphics[width=.9\linewidth]{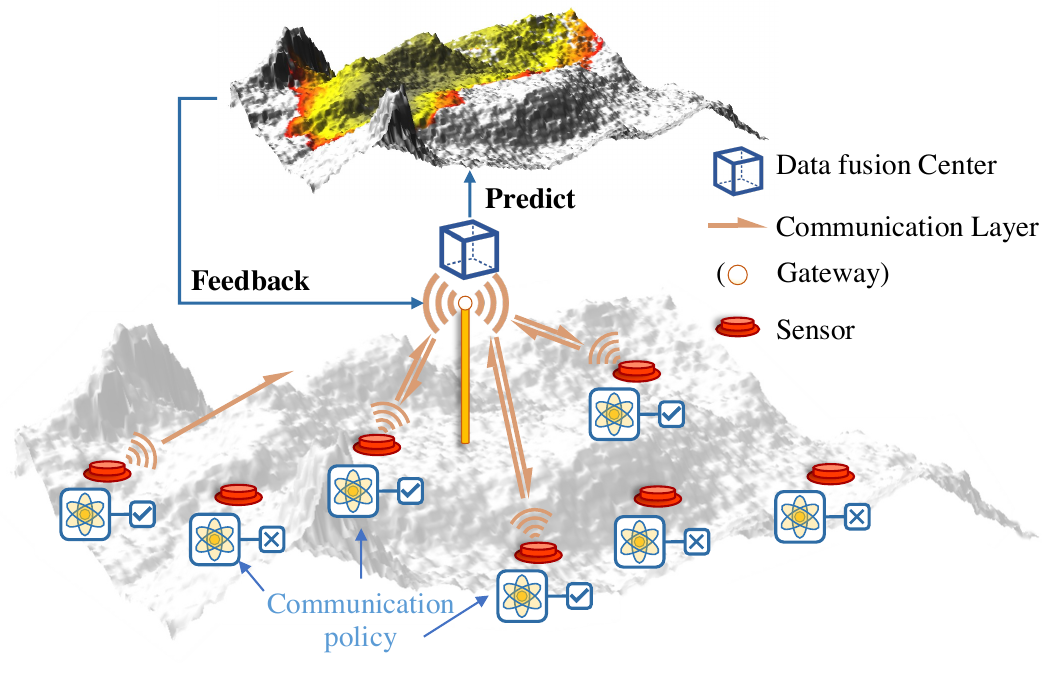}
    \caption{System design for wildfire tracking. Multiple IoT sensors periodically monitor the environment and send sensed data to a central gateway over LPWAN. The gateway uses its received data to predict future wildfire spread.}
    \label{fig:system}
\end{figure}

The rise of the Internet of Things (IoT) with ubiquitous sensing capabilities~\cite{sehrawat2019smart} enables numerous services and data-driven applications in domains ranging from healthcare to civil engineering, by interconnecting a large number of devices equipped with networking, sensing, and processing capacities~\cite{iot}. 
In this work, we study the use of IoT sensors for environmental tracking, where each sensor monitors the environmental conditions at its location and decides on whether to forward its locally collected data to a central location that aims to track the overall environmental conditions as accurately as possible.
A typical example is the wildfire tracking problem illustrated in Figure~\ref{fig:system}. The goal is to predict the spread of a wildfire by aggregating environmental data (e.g., wind direction, temperature) monitored by IoT sensors. The central location, i.e., a firefighting command center, can then use these predictions to deploy firefighting resources or issue evacuation orders.
We consider centralized data aggregation, since distributed prediction methods may require more communication to achieve the same prediction accuracy.

In practice, effective IoT sensing for environmental tracking requires non-trivial \textbf{communication planning and coordination}, mainly due to sensors' power and bandwidth constraints. 
Most IoT devices have \emph{limited power supply}, since they are battery-powered or do not have access to mains power. 
Since turning on a radio to send data to a gateway requires significant power~\cite{orsino2016energy}, asking each sensor to continually send data will then quickly drain their limited power supply. Less frequent data reports may lead to lower tracking accuracy, but not always: environmental data in real-world applications usually exhibit spatiotemporal correlation, which can make some sensors' data redundant. Thus, carefully designed communication scheduling is needed to balance the tradeoff between sensors' power consumption and tracking accuracy.

This tradeoff is further complicated by outdoor IoT devices' use of long-range low-power wide-area networking (LPWAN) technologies with stringent \emph{bandwidth constraints} (e.g., LoRa, SigFox)~\cite{lora_survey} to transmit their data to gateways. 
Multiple sensors' simultaneous transmissions, then, may cause communication failures due to channel congestion and interference. Even worse, IoT sensors are usually deployed in an ad hoc manner, so their exact interference range, which depends on their physical locations, is not known in advance. Effective communication coordination should then learn these failure patterns and account for their effect on the tracking accuracy.

\subsection{Motivating Example: Wildfire Tracking}\label{sec:wildfire}
We illustrate the importance of communication planning and coordination that addresses the above concerns by considering the example problem of \textbf{wildfire tracking}. Large wildfires cause severe air pollution, burn millions of acres, and cost billions to suppress~\cite{ca_wildfire_stats}. To fight wildfires and reduce damage, fire chiefs rely on real-time environmental data to track and predict the wildfire spread.  Although satellite imagery has been extensively used for wildfire detection, its lengthy scanning cycles and poor resolution have driven a shift towards wireless sensor networks for time-critical wildfire prediction and detection~\cite{KAUR2019171,iot-fire,aeris}. 
Combined with physics-based wildfire simulators, networks of IoT devices deployed over the anticipated burn area can collaboratively collect wildfire-relevant data and send it to a fire command center in time.

Ensuring that IoT sensor deployments collect useful wildfire data, however, requires sensor coordination. For example, if an anemometer on the fire boundary reports east wind, we may want sensors to the west of it to report next because the fire is likely to propagate west. 
Ignoring such spatiotemporal correlation in the utility of collected data, e.g., by asking each sensor to continually send data, will quickly drain IoT sensors' limited power supply. Wildfires can burn for days or weeks, so sensors' batteries should last that long. Without coordination, sensors can also easily overwhelm the wireless channel: once the fire arrives at a new location, all nearby sensors may tend to report their newly observed data simultaneously, which leads to severe channel collisions. Coordination without requiring explicit sensor communication is then needed to enable efficient tracking with fewer communication failures.

\subsection{Research Challenges}
The power constraints of IoT sensors and the bandwidth limitations of communication channels make it impractical to deploy centralized communication policies. Thus, the goal of this paper is to find \emph{distributed} policies for IoT sensors to maximize the overall tracking accuracy, subject to these limitations. We summarize three main challenges of doing so.

    (1) {\textbf{Power and bandwidth constraints}} make it hard to find the policy that maximizes the overall tracking accuracy. The power constraints limit the amount of data that can be shared by the sensors, while the bandwidth constraints depend on the real-world wireless environment and thus might not have explicit models. Finding the optimal policy becomes even more challenging if we aim to maximize the average tracking accuracy over time instead of the instantaneous accuracy, as the power and bandwidth constraints can change over time and the myopically optimal policy might not be optimal overall.
    
     (2) {\textbf{Spatiotemporal correlation}}. 
    Spatiotemporal correlation in the environment makes it possible to use collected data to infer the environmental conditions at other locations, while such correlation in wireless interference could lead to additional communication failures. As a result, reporting redundant data would not help improve the tracking accuracy but incur additional power consumption. However, since IoT networks are usually ad hoc and the environmental and wireless conditions change over time, we would not know the spatiotemporal correlation as a priori and must learn it instead.
    
     (3) {\textbf{Importance of local observations}}. Since the sensors will make individual transmission decisions based on their local information, it is difficult for them to directly optimize the overall tracking accuracy, which depends on the data reported by all sensors. Instead, each sensor should understand how its local observations would contribute to the global tracking problem, which is complicated by spatiotemporal correlation in the data: for example, the presence of a wildfire at one sensor might be inferred by the data of wildfire at a nearby sensor. Thus, we must find an appropriate metric to evaluate the importance of each sensor's local observation. This metric would serve as the backbone of the design for any intelligent distributed communication policy.



\subsection{Our Contributions}
In this work, we formulate the communication planning problem for general environmental IoT sensing with constraints. 
After reviewing related literature (Section~\ref{sec:related}), we make the following contributions:
    
    \textbf{\textit{Problem formulation}} (Section~\ref{sec:formulation}). We combine the energy cost of communication, tracking accuracy, and models for successful data transmission into our formulation, aiming at balancing the tradeoff between conserving device power and maintaining high tracking accuracy. The tradeoff can be customized to value different factors, e.g., if some sensors are less power-hungry than others. 
    We then propose a \emph{metric} to quantify the value of each sensor's local observations, which helps connect local decisions to the global tracking accuracy.
    
    \textbf{\textit{Multi-agent RL}} (Section~\ref{sec:planning}). We use an MDP (Markov Decision Process) to model the multi-round communication planning and propose multi-agent RL (reinforcement learning) algorithms to learn efficient communication policies. The compute-intensive RL policy training can be done offline in a simulated environment so that the IoT sensors only need to run those learned policies at the deployment stage, which does not require much compute resource. Multi-agent RL (MARL) would also learn and exploit the spatial-temporal correlation implicitly, based on our defined data values.
    
    \textbf{\textit{Theoretical analyses}} (Section~\ref{sec:analysis}). We provide theoretical analyses of the optimal policies given full information under simplified communication models, with an example of tracking an environment with binary states (e.g., ``active fire/inactive''). These analyses allow us to design heuristics inspired by the optimal policies in this simplified setting. 
    
    \textbf{\textit{Experimental results}} (Section~\ref{sec:experiments}). We demonstrate that our method outperforms baselines, including the heuristics from our theoretical analysis, and effectively balances the tradeoff between tracking accuracy and communication cost.

We finally conclude in Section~\ref{sec:conclusion}.

\section{Related Work}\label{sec:related}
\textbf{IoT sensing} has been extensively studied, and our work is related to the IoT data collection problem~\cite{luong2016data,orsino2016energy,plageras2018efficient,liu2018blockchain,wu2019learn}. 
\cite{orsino2016energy} proposed an energy-efficient data collection scheme for IoT in a smart city scenario, based on device-to-device communications. \cite{plageras2018efficient} proposed new methods to collect and manage sensor data in a smart building operating in the IoT environment. RL is used in \cite{liu2018blockchain} with blockchain to maximize the amount of collected data while ensuring secure and reliable data sharing in Industrial IoT, and in~\cite{wu2019learn} for adaptive sampling in field reconstruction, proposing to determine the most informative sensing location and thus reducing the communication cost. However, their data collection frameworks only choose one node to collect data at each timestep, while our framework allows more complicated coordination of multiple nodes. Another line of research studies \textbf{mobile crowdsensing}~\cite{capponi2019survey} for environmental monitoring, e.g., air quality~\cite{wu2020sharing}, noise~\cite{alsina2016design}, temperature~\cite{koukoutsidis2017estimating}, etc. monitoring. However, these works leverage mobile devices, whose wireless environments are difficult to learn due to mobility, to collectively sense and collect data, while we design communication policies for IoT sensors deployed at fixed locations in an ad hoc manner.

\textbf{Multi-agent reinforcement learning} has been proposed to solve many types of problems in IoT networks, from optimal routing to power control~\cite{li2022applications,khan2012resource}. However, few of these current applications consider application-layer objectives like the usefulness of spatiotemporal sensor information, as we do here. A notable exception is UAV trajectory optimization~\cite{hu2020cooperative}, which attempts to control UAV movement instead of sensor communication decisions to maximize UAV communication ability. MARL-based communication protocols often consider a device-to-device topology and do not take advantage of a base station's ability to help coordinate users~\cite{nisioti2019robust}. Other works focus on using RL to learn MAC protocols, with the goal of saving energy~\cite{savaglio2019lightweight,sahraoui2022schedule} or learning to handle a specific wireless medium~\cite{zhao2020reinforcement}.
Although there exist many RL applications in networked systems~\cite{lei2020deep}, we are unaware of others that apply multi-agent RL~\cite{zhang2020marl} to environmental sensing in IoT networks with power and bandwidth constraints.

Prior MARL work on effective communication coordination requires centralized orchestration~\cite{sukhbaatar2016learning}, which violates IoT systems' stringent resource constraints. Few existing MARL approaches consider realistic communication constraints~\cite{kim2019learning} or try to lower communication overhead~\cite{zhang2019efficient}. Prior work mainly focuses on adding extra mechanisms (e.g., reducing the size of messages, introducing penalty terms to restrict communication behaviors) to existing MARL methods~\cite{marl_survey}. Most existing IoT communication protocols, on the other hand, optimize for generic energy expenditure or throughput objectives~\cite{li2022applications}, in contrast to our dedicated sensing objective.  


\section{Problem Formulation}\label{sec:formulation}
{
We consider a wireless network of $N$ sensors that monitor the physical conditions of the environment at each sensor's location. Let the sensors sample the environment at discrete time steps $t=1,2,\cdots$. At each time step, each sensor makes an independent decision on whether to send its locally collected data to a central location. The goal is for this central location to track the environmental conditions as accurately as possible over time, subject to power and channel limitations. 

\subsection{Tracking Environmental Conditions}
We assume all sensors can monitor the same set of physical conditions, e.g., an indicator of a wildfire's presence, humidity, or temperature, at the sensor's location. As the environment changes over time, the $i$-th sensor can collect the real local physical conditions at time $t$ as $x^i_t$. {Due to memory constraints at the sensor, we assume the sensed data is deleted from the sensor, whether transmitted or not, after each timestep.} We denote the real conditions of the whole area as $\textbf{X}_t=\{x^i_t|i=1,...,N\}$. Each sensor can send its collected data $x^i_t$ to a central location, which we call the \emph{information center}, that tracks the environmental conditions of the whole area. The information center holds \emph{beliefs} $\textbf{Y}_t=\{y^i_t|i=1,...,N\}$ of the local conditions at each sensor location $i$ at time $t$, based on previously collected data {(i.e., data received by the information center through time $t - 1$)}.

We want to track the physical conditions of the environment as accurately as possible. We define an \textbf{error loss function} $e^i_t=\bar{l}(x^i_t-y^i_t)$ for each sensor location $i$, and the error loss at each time step $t$ is 
\begin{equation*}
    L_t = \sum_{i=1}^N e^i_t= \sum_{i=1}^N\bar{l}(x^i_t-y^i_t)=l(\textbf{X}_t-\textbf{Y}_t).
\end{equation*}
The increase in the error loss  can be linearly approximated as
\begin{align}\label{eq:loss_diff}
L_{t+1}-L_t & \approx  \nabla l\left({\mathbf{X}_t}-\mathbf{Y}_t\right)\bigl(\Delta_t\mathbf{X}_t-{\Delta_t \mathbf{Y}_t}\bigr),
\end{align}
where $\Delta_t\mathbf{X} = \mathbf{X}_{t+1} - \mathbf{X}_{t}$ and $\Delta_t\mathbf{Y} = \mathbf{Y}_{t+1} - \mathbf{Y}_t$. 
Therefore, the increase in the error loss depends on the difference between the change in the physical conditions of the environment $\Delta_t\mathbf{X}$, and the change in the belief $\Delta_t\mathbf{Y}$. Intuitively, we want to minimize the discrepancy between how the environment actually evolves and how we update our beliefs. Note that $\Delta_t\mathbf{Y}$ depends on the belief system as well as the sensor actions (i.e., the data received at the information center).
\subsection{Energy Cost and Communication Limitation} \label{sec:comm_formulation}
 We consider possibly hundreds of sensor nodes with limited power supply sharing  $M$ wireless low-bandwidth channels. 
 {If the central location received data from every sensor in every time step, its belief could exactly track the actual state of the environment. However, sending data expends limited sensor resources, and each wireless channel has a finite transmission capacity. We model these limitations in this section.} First, sensors consume energy when turning on the radio and sending a data packet. {Intuitively, sensors should only spend the resources to transmit data if doing so would meaningfully reduce the error loss.} We consider a fixed energy cost of $c^i$ for each transmission attempt made by sensor $i$. The \textbf{total energy cost} of transmission $I_t$ (the set of sensors that
transmit) is 
\begin{equation}
    c(I_t)=\sum_{i\in I_t}c^i.
    \label{eq:cost}
\end{equation}

These wireless channels for outdoor sensing have stringent bandwidth constraints. To avoid severe channel congestion and save device power, only a  subset of sensors report at each time step. 
Some of their data transmissions may fail due to channel noise and congestion, and we denote the set of sensors that transmit successfully as $I^s_t$ and $I^s_t\subseteq I_t\subseteq [N]$.

 The total throughput depends on the channel bandwidth. For one channel, given $n$ transmissions from sensors (assuming all data packets are of the same length), we discuss the \textbf{probability of transmission success} for two models:
\begin{itemize}
    \item \textbf{Bandwidth-limited channel (PHY):} At the physical layer, by the Shannon-Hartley theorem,  the channel capacity is approximately linear in bandwidth when the signal-noise ratio is large.  
    We thus consider the maximum channel capacity in units of packets per time step to be $C$, yielding the probability of transmission success: 
    \begin{equation}\label{eq:comm_bw}
        P_{bw}(n) = \begin{cases}1-\epsilon, & n\le C \\ C/n-\epsilon, & n > C,\end{cases}
    \end{equation}
     where $\epsilon$ is the failure rate caused by background noise. 
    
    \item \textbf{ALOHA random access channel (MAC):} Due to the lack of centralized control for outdoor IoT sensing, it is natural to use random access protocols at the MAC layer~\cite{sinha2017survey}. We consider a simple ALOHA system, where all devices randomly access the channel whenever they have data to send or after a random waiting time. We assume sensors will randomly start transmission in a fixed time window $W$ at each time step. The resulting probability of transmission success is 
    \begin{equation}\label{eq:comm_a}
        P_a(n)=e^{-2n/A}
    \end{equation}
    where $A=W/\tau$ is the ratio of the transmission time window $W$ to a packet transmission time $\tau$. The maximum throughput is $0.5A/e$, reached when $n=A/2$.
\end{itemize}
{We consider these two models in Section \ref{sec:optimal}'s analysis and use a realistic LoRa simulator in Section~\ref{sec:experiments}'s experiments.}


\subsection{Communication Planning under Constraints} 

{Sensors make their transmission decisions by balancing the relative ``value'' of their data for reducing the error loss, communication cost, and available bandwidth. To formalize this process,} we use a general value function $v(\mathcal{C})$ to define the data value of a set of data $\mathcal{C}$ as it contributes to updating $\textbf{Y}(t)$ such that maximizing the sum of the data value is equivalent to minimizing the error loss (see Section \ref{sec:data-value}). Over a time horizon $T$, we define the {sensors' collective} objective function as the sum over the value of the collected data at the expense of the (energy) cost of communication \eqref{eq:cost}

\begin{equation}\label{eq:obj}
    \max_{I_t} \sum_{t=1}^T \left(v(X^s_t) - wc(I_t)\right),
\end{equation}
where $X^s_{t}=\{x^i_t,\forall i\in I^s_{t}\}$ is the set of data collected at $t$ and $w$ is a user-tunable weight indicating how we much want to trade off between conserving device power and maintaining high tracking accuracy.  A greater $w$  will cause the system to be more conservative in energy expenditure.

For simplicity, we estimate the data value of a set as the sum of the data values of individual sensors' contributions:
\begin{equation}
    v(X^s_{t}) = \sum_{i\in I^s_t}v^i_t.
\end{equation}

We consider stochastic policies $\pi(t)$ where each sensor $i$ will send locally collected data with probability $p_i(t)$, and the probability of success is $P_i(\mathcal{C})$ given a set $\mathcal{C}$ of sensors decide to transmit. Consider the expected value of $g(t)\equiv v(X^s_{t}) - wc(I_t)$ at time $t$:
\begin{equation}\label{eq:objective_step}
    \mathbf{E}[g(t)] =\sum_{I_t\sim \pi}f(I_t)\left(\sum_{i\in I_t}P_i(I_t)v^i_t - wc^i\right),
\end{equation}
where $f(I_t)=\prod_{i\in I_t}p_i(t)\prod_{i\in[N]-I_t}(1-p_i(t))$ is the probability of the set $I_t$ decide to send based on the policy $\pi$.

\subsection{Defining the Data Value}\label{sec:data-value}
The data value assigned to each data sample should reflect its contribution to reducing the error loss. We assume the latest belief $\mathbf{Y}_t$ is built upon all collected data $\mathcal{X}_t=\bigcup_{t'\le t} X^s_{t'}$ by some function $\mathbf{h}$. Mathematically, $\mathbf{Y}_t = \mathbf{h}(t|\mathcal{X}_t)$.

We consider the belief system  updated step-wise, i.e., $\mathbf{Y}_{t-1}\rightarrow\mathbf{Y}_t\rightarrow\mathbf{Y}_{t+1}\rightarrow \cdots$ (Figure \ref{fig:framework}). To transit from step $t$ to $t+1$, the belief system will (i) incorporate newly collected data $X^s_{t}$ and (ii) generate new belief $\mathbf{Y}_{t+1}$. Since channel limitations prevent collecting all sensors' collected data at time $t$, 
$\mathbf{h}$ fills in the missing data values. 
Defining $\mathbf{h}$ is application-specific, e.g., using knowledge of spatial correlation in the environment's physical conditions. 

$\mathbf{h}(\cdot|\mathcal{X})$ is a function of time based on the set of data $\mathcal{X}$. We use $\mathbf{Y}^{\mathcal{X}}(\cdot) =  \mathbf{h}(\cdot|\mathcal{X})$ to denote the belief function if $\mathcal{X}$ has been collected and $\mathbf{Y}_t = \mathbf{Y}^{\mathcal{X}_t}(t)$. Consider $\mathbf{Y}^{\mathcal{X}_{t-1}}(t)=\mathbf{h}(t|\mathcal{X}_{t-1})$, which is the belief at time $t$ based on ${\mathcal{X}_{t-1}}$ and demonstrates how well $\mathbf{h}$ approximates the real physical conditions without knowledge from time $t$. We define a step-wise model loss $L^m_t$ as the error loss of  $\mathbf{Y}^{\mathcal{X}_{t-1}}(t)$, 
\begin{equation}
    L^m_t = l(\mathbf{X}_t-\mathbf{Y}^{\mathcal{X}_{t-1}}(t))=l(\mathbf{X}_t-\mathbf{h}(t|\mathcal{X}_{t-1})).
\end{equation}
$L^m_t$ can be used as a baseline for evaluating the data value of additional data collected at time $t$, and 
\begin{equation}\label{eq:g0}
    L^m_t =l(\mathbf{X}_t-\mathbf{Y}^{\mathcal{X}_{t-1}}(t))\ge l(\mathbf{X}_t- \mathbf{Y}_t)= L_t,
\end{equation}
 assuming that it is always beneficial to incorporate more data. 

We define the \textbf{data value} $v(\mathcal{C}|\mathcal{X})$ of a set of data $\mathcal{C}$ as the reduction in the error loss at time $t$ if $\mathcal{X}$ has been collected before $t$ and only $\mathcal{C}$ is collected at time $t$. 
\begin{align}
    v(\mathcal{C}|\mathcal{X}_{t-1})&=l(\mathbf{X}_t- \mathbf{Y}^{\mathcal{X}_{t-1}}(t)) - l(\mathbf{X}_t- \mathbf{Y}^{\mathcal{X}_{t-1}\cup \mathcal{C}}(t))\nonumber\\
    & = L^m_t - l(\mathbf{X}_t- \mathbf{Y}^{\mathcal{X}_{t-1}\cup \mathcal{C}}(t)).
\end{align}
The data value of a single data sample $x^i_t$ can then be defined as $v^i_t = v(\{x^i_t\})$.
Given a set of data $X^s_t$ collected at time $t$, we have $\mathbf{Y}^{\mathcal{X}_{t-1}\cup X^s_t}(t) = \mathbf{Y}^{\mathcal{X}_{t}}(t)=\mathbf{Y}_t$. Therefore, 
\begin{equation*}
        v(X^s_t|\mathcal{X}_{t-1}) = L^m_t - l(\mathbf{X}_t- \mathbf{Y}_t) = L^m_t - L_t.\label{eq:g1}
\end{equation*}
We approximate the total data value as the sum of the individual values, facilitating Section~\ref{sec:mdp}'s MDP formulation:
\begin{equation}
    v(X^s_t)\approx \sum_{i\in I^s_t} v^i_t = \sum_{i\in I^s_t} v(\{x^i_t\}).
\end{equation}


\section{Communication Planning}\label{sec:planning}

\begin{figure}
    \centering
    \includegraphics[width=.7\linewidth]{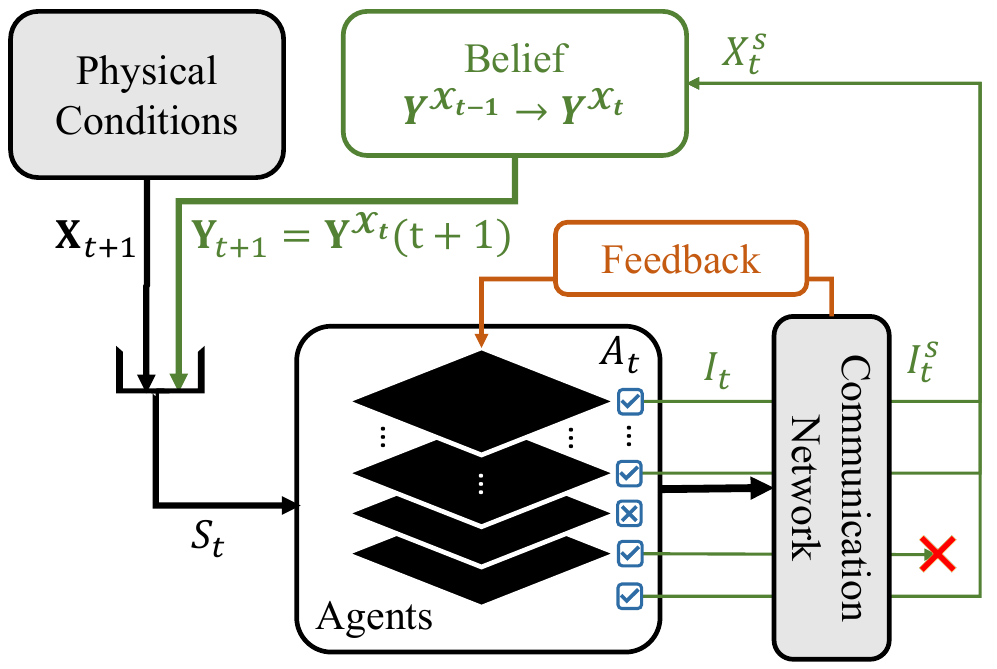}
    \caption{EnvSen Framework for sensors' communication decisions.}
    \label{fig:framework}
\end{figure}

Planning the communication for efficient data collection is  challenging. Firstly,  evaluating the value of collecting a data sample can still be highly complex, depending on how the belief builds on the collected data. Moreover, the wireless communication channel is in practice highly complex, and the probability of transmission success $P_i$ for a sensor $i$ can be affected by the noise from the environment and  the nearby sensors. Finally, finding the optimal solution to $\min_{I_t}\sum_{i\in I_t} P_i(I_t)v^i_t$, which does not even include communication costs, is NP-hard, as can be shown by reduction to the knapsack problem. We thus introduce an MDP formulation to maximize \eqref{eq:objective_step} over time and a corresponding MARL solution framework, which we evaluate in Section~\ref{sec:experiments}.
\subsection{MDP Formulation}\label{sec:mdp}
We model the multi-round communication planning of the sensors as a Markov decision process (MDP). We consider a multi-agent formulation, where each sensor is an agent. We define the state, action, and reward as follows.
\paragraph{State} $S_t = (\mathbf{X}_t, \mathbf{Y}_t)$. We also define $s^{i}_{t} = (x^{i}_{t}, \mathbf{Y}_t)$ as the local state of agent $i$.
\paragraph{Action} $A_t = \{a^i_t\}_{i=1}^N$, where $a^i_t \in \{0,1\}$ is the local action of agent $i$ and $= 1$ if sensor $i$ sends its local data.
\paragraph{Reward}  $R_t = \sum_{i} a^i_t(P_i(A_t) v^{i}_t - wc^i)$.
We also define $r^{i}_{t} = a^i_t(P_i(A_t) v^{i}_t - wc^i)$ as the local reward of agent $i$.
Note that we slightly abuse the notation of $P_i(\cdot)$ to represent the successful transmission probability of agent $i$ under action $A_t$.

At each round $t$, each agent $i$ observes its current local state $s^{i}_{t}$ from the environment, then chooses an action $a^{i}_{t}$; the environment moves to a new global state $S_{t+1}$ based on the global action $A_t$, and generates a global reward $R_t$; the agent then chooses the next action $a^{i}_{t+1}$ based on $s^{i}_{t+1}$ and repeat the interaction with the environment. The goal of the learning agent is to learn a policy $\pi_i: s^i_t \rightarrow a^i_t$ that maximizes the expected cumulative reward. Note that the state of agent $i$ includes $\mathbf{Y}_t$, the current global belief at time $t$. In practice, since the beliefs $\mathbf{Y}$ are calculated at the central gateway and not the individual sensors, the gateway may need to broadcast these beliefs to each sensor in order for the sensors to make individual decisions that solve the MDP. These can be low-power broadcasts~\cite{daskalakis2017ambient}, or if the beliefs do not account for spatiotemporal correlations, e.g., due to a lack of models for such correlation, the sensors may be able to estimate the beliefs based on their past data transmissions to the gateway.

\subsection{MARL Solution}
Based on the MDP formulation, we propose a MARL-based Environmental Sensing (EnvSen) framework, as illustrated in Fig. \ref{fig:framework}. Traditional single-agent RL that considers the whole IoT network as a single agent would suffer the issue of scalability (state and action space could be exponentially large), while MARL treats each sensor as an agent so that they can learn their own policies based on their local states and feedback individually. Since we do not know the communication network $P_i(A_t)$  a priori or the action of other sensors, we use $\bar{r}^i_t=a^i_t(\mathbf{1}^i_t v^{i}_t - wc^i)$ where $\mathbf{1}^i_t$ indicating whether the transmission is successful or not. This allows each agent $i$ to get an unbiased estimate of the true reward and optimize the local policy $\pi_i$ by maximizing the local reward $r^i_t$. 

EnvSen has two aspects:  (1) The local reward of agent $i$ includes the data value $v^i_t$ that encodes the  spatiotemporal correlations in the utility of collected data. This can help guide the agent to learn the optimal policy that utilizes such correlations. (2) We allow sensors to leverage feedback of the network conditions from the gateway (e.g., how busy the current network is).  

\subsubsection{Wildfire Tracking}
We use the wildfire tracking problem to illustrate the MARL method. As discussed in Sec.~\ref{sec:wildfire}, the goal is to track the spread of a wildfire with distributed IoT sensors.
We consider binary fire state $x^i_t \in \{0,1\}$ for each sensor $i$ and realistic fire propagation where the changes in fire states have spatiotemporal correlation. 
It is also possible to design a more sophisticated belief system that utilizes our knowledge of the fire propagation process (e.g., the effect of wind direction and speed on the fire propagation). To track the wildfire, in addition to the simple binary indicator of fire state, the sensor may also collect additional environmental data, such as the wind speed and direction,  which may help make the belief function more accurate. Regardless, we can define the data value given the belief function $\mathbf{h}$.

\section{Analysis of Simplified Settings}\label{sec:analysis}
We next consider a simplified communication network model where the transmission success probabilities of  all the sensors are the same and depend on the number of active sensors only, i.e., $P_i(\mathcal{C})=P(|\mathcal{C}|),\forall i\in\mathcal{C}$. $P(n)$ is monotonically decreasing, e.g.,  \eqref{eq:comm_bw} and \eqref{eq:comm_a}. In this case, the step-wise objective \eqref{eq:objective_step} becomes:
\begin{equation}\label{eq:obj_s}
    \mathbf{E}[g^s(t)] =\sum_{I_t\sim \pi}f(I_t)\left(P(|I_t|)\sum_{i\in I_t}v^i_t - w\sum_{i\in I_t}c^i\right).
\end{equation}
 We show that optimizing \eqref{eq:obj_s} is nontrivial even under the simplified network model and full information at all sensors. We then analyze the special case of binary tracking (e.g., of wildfires), using our analysis to define policy baselines used in Section~\ref{sec:experiments}'s experiments.

\subsection{Optimizing with Full Information}\label{sec:optimal}
We analyze the optimal solution (i.e., maximizing (\ref{eq:obj_s}) when all sensors know the real physical conditions $\mathbf{X}_t$, the belief function $\mathbf{h}$, the value function $v$, and the channel model $P(n)$. Since in practice, full information is not available, the analysis provides an  upper bound for the performance. 


\subsubsection{Centralized Policy}\label{sec:centralized-optimal} We first consider a centralized deterministic policy $\pi_c$ that will determine the actions for all sensors, i.e.,  $\mathbf{a}_t=\{a^i_t\in \{0,1\}, \forall i\in [N]\}$, which upper bounds the performance of distributed communication policies.
Let $I^a_t=\{i|a^i_t=1\}$ denote the set of transmitting sensors. The optimal centralized policy maximizes \eqref{eq:objective_step},
\begin{equation*}
    I^{a^*}_t = \argmax_{\substack{I^a_t}} \sum_{i\in I^a_t}P(|I^a_t|)v^i_t - wc^i= \argmax_{\substack{I^a_t}} \sum_{i\in I^a_t}g_i(|I^a_t|)
\end{equation*}
where $ I^{a^*}_t$ is the optimal set and $g_i(n)=P(n)v^i_t-wc^i$. 
We can find the optimal solution by the following procedure (Algorithm~\ref{alg:optimal}): for $n=0,1,2,...$, pick the top $n$ sensors after sorting by $g_i(n)$, and find the sum, and stop incrementing $n$ until the sum does not increase.
\begin{algorithm}[t]
\begin{algorithmic}
\STATE $g_m\gets 0,I^*_t \gets \emptyset$
\STATE $g\gets 0,n\gets 1$
\WHILE{$g>=g_m$}
\STATE $I_t$: sort $i$ by $g_i(n)=P(n)v^i_t-wc_i$
\STATE $I_t\gets I_t[:n]$: select top $n$ \COMMENT{Greedily choose $n$ sensors}
\STATE $g\gets \sum_{i\in I_t}g_i(n)$
\IF{$g>=g_m$}
\STATE $g_m\gets g, n\gets n+1$ \COMMENT{Stop if \eqref{eq:obj_s} is decreasing}
\STATE $I^*_t \gets I_t$
\ENDIF
\ENDWHILE
\end{algorithmic}
\caption{Centralized communication policy}
\label{alg:optimal}
\end{algorithm}
\begin{proposition}\label{prop:optimal}
Algorithm~\ref{alg:optimal} yields the optimal communication policy if all sensors have the same communication cost $wc^i$.
\end{proposition}
\begin{IEEEproof}
Suppose for a given number of transmitting sensors $n$, we order the sensors in decreasing order of $g_i(n)$. Then the order of the sensors does not change with $n$, and the optimal solution is to greedily choose sensors in this order until their sum $\sum_i g_i(n)$ stops increasing (Algorithm~\ref{alg:optimal}).
\end{IEEEproof}

When all sensors have the same data values $v^i_t$, the optimal policy is further simplified:
\begin{corollary}
The optimal policy finds the optimal number of sensors to report:
$n^* = \argmax_{n\in \mathbb{N}_{0}, n\le N}n(vP(n)-wc)$.
Any subset of $n^*$ sensors is equivalently optimal. 
\end{corollary}
\begin{IEEEproof}
    Follows directly from Proposition~\ref{prop:optimal}.
\end{IEEEproof}
For example, consider the bandwidth-limited model with capacity $C<N$ in \eqref{eq:comm_bw}. 
The optimal centralized policy is to either use the full communication capacity $(n^* = C)$ or to select $n^*$ such that if $n > n^*$ sensors transmit, the cost $wc^i$ exceeds the data value $P(n)v^i$. 
Similarly, the optimal centralized policy for the ALOHA model \eqref{eq:comm_a} is to stop communication if $wc/v\ge e^{-1}$, and $n^*=\lfloor A/2\rfloor$ if  $wc/v<e^{-1}$.  
%
%
%

%
We use Algorithm~\ref{alg:optimal} as the basis for our subsequent analysis. 

\subsubsection{Distributed Execution} We now consider sensors that independently decide whether to forward the data or not, without knowing the actions of any other sensors. 
%
With full information, we see from Algorithm~\ref{alg:optimal} that all sensors can find the same optimal number of transmissions $n^*$; a sensor will send the  data if its value of $g_i(n^*)$ is among top $n^*$. 
However, this policy is difficult to distributedly execute when multiple sensors have the same  $g_i(n)$ value. Centralized execution allows selecting a subset of such sensors, but in the distributed case, sensors cannot coordinate to do so.  Since two sensors under the same state will take the same action, a naive deterministic policy that results in all of them sending or not sending at the same time can hardly be optimal. 

Instead, we can use a stochastic policy that outputs a probability of sending data. First, suppose we find the top $k \le n$ sensors with the highest $g_i(n)$ values, where $k$ is the largest number such that $g_i(n)$ of sensor $k$ is greater than that of the remaining $n-k$ sensors. For the remaining $n-k$ sensors, we have a set of $m$ sensors with the same data value and communication cost, and $m>n-k>0$. Denote the set of $m$ sensors as $I^m$ and their $g_i(n)=\phi(n)$ for all $i\in I^m$. Assume they follow the same stochastic policy $\pi^m$ that lets them each transmit with probability $p$. The expected sum of rewards is:
\begin{align}
    & \sum_{\substack{I_t\sim \pi_m, \\I_t\subseteq I^m}} f(I_t)\sum_{i\in I_t}\left(P(k+|I_t|)v^i_t - wc^i\right)\nonumber\\
    =& \sum_{j=1}^m f(j;m,p)j\phi(k+j)\label{eq:1}
\end{align}
where $f(j;m,p)=\binom{m}{j}p^j(1-p)^{m-j}$ is the binomial distribution of getting $j$ successes in $m$ independent Bernoulli trials with success probability $p$.  $k+|I_t|=k+j$ is the resulting total number of transmissions. 

\begin{proposition}\label{prop:max_return}
The expected total reward is maximized for a given $n$ by, in addition to selecting the top $k$ sensors with the highest $g_i(n)$ values, having  each sensor in $I^m$ to report with a probability that maximizes Eq.~\eqref{eq:1}.
\end{proposition}

\begin{corollary}
When all the sensors have the same data value and energy cost, i.e., $v^1_t=v^2_t=...=v^N_t=v, c^1=c^2=...=c^N=c$, Proposition~\ref{prop:max_return} reduces to finding the optimal transmission probability:
\begin{align}\label{eq3}
    p^* = \argmax_{p\in[0,1]} \sum_{j=1}^N \binom{N}{j}p^j(1-p)^{N-j} j(vP(j)-wc),
\end{align}
\end{corollary}
Given full information, we can find the $n^*$ and  $p^*$ exactly. We consider the two simple communication channel models in Section \ref{sec:comm_formulation}. For the bandwidth-limited model with capacity $C<N$ and minimum failure rate $\epsilon$, $n^*=0$ if  $wc/v>=1-\epsilon$, and $n^*=C$ otherwise. 
%
For the stochastic policy, the optimal transmission probability can be found by finding a real root $\in[0,1]$ after taking the derivative of \eqref{eq3}. For the bandwidth-limited channel, $p^*\approx C/N$ when $wc/v\approx 0.5$.

\subsection{Binary Tracking} 
Finally, we use our results from Section~\ref{sec:optimal} to discuss the special case where $x^i_t$ is a binary indicator of some event, e.g., whether sensor $i$'s location is on fire at time $t$ for the wildfire tracking problem, that takes value 0 or 1. 


Let $z^i_t$ be the belief from $\mathbf{Y}^{\mathcal{X}_{t-1}}(t)$, i.e., the belief if $x^i_t$ is not reported. There are two cases: $x^i_t=z^i_t$ and $x^i_t\neq  z^i_t$. If they are equal, the data value $v^i_t$ of collecting $x^i_t$ is zero because it provides no additional information about location $i$ or any other location. If they are different, we have a positive  data value  by correcting the belief  $y^i_t=z^i_t\rightarrow y^i_t=x^i_t$. Assume all locations are of equal importance and have the same data value $v$ if a correction is made. We denote the set with positive data value to be $G_t$, i.e., $G_t=\{i|\forall i \in [N], x^i_t\neq z^i_t\}$.

We can then characterize the optimal policy given the full information by directly applying Algorithm~\ref{alg:optimal}:
\begin{corollary}
The optimal communication policy is for only sensors in $G_t$ to report.
\end{corollary}
Communication from any sensor not in $G_t$ will only introduce additional cost without gaining any data value. 


As discussed in Section~\ref{sec:optimal} above, implementing the centralized optimal policy is difficult when sensors make independent decisions, especially when sensors cannot estimate their data values, as they do not necessarily know the gateway's beliefs $y^i_t$ (see Section~\ref{sec:mdp}'s discussion). 
If sensors can estimate the data value, e.g., the belief $y^i_t$ uses only information from sensor $i$, we can simply add a rule in the distributed policy to suppress sensor communication when $z^i_t=x^i_t$. We thus reduce redundant communication and mitigate channel congestion; this insight is the basis for our heuristic benchmark algorithm in Section~\ref{sec:experiments}'s experiments.




\section{Experiments}\label{sec:experiments}
We evaluate our approach for wildfire tracking. The goal is to track the wildfire as accurately as possible under resource constraints. We consider simple binary fire states indicating whether there is active wildfire at each sensor location. The belief system is built on a simple propagation model that estimates the fire propagation given limited sensor data. The sensors, in addition to monitoring the local fire state, also report wind direction and speed that can make the prediction of the fire propagation more accurate. We use the wildfire modeling tool provided by GRASS GIS~\cite{neteler2012grass} to generate $200$ instances of realistic wildfire propagation data, with different starting locations and wind conditions, as the ground truth. 

\subsection{Experimental Setup}
We set up 200 sensors at random locations and simulate two types of communication networks: a simple bandwidth-limited channel and a realistic LoRa wireless network.  LoRa is one of the most energy-efficient communication technologies for the IoT, which enables long-range and bidirectional communication between low-power devices and always-on gateways~\cite{iot_lora,lora-popularity}. For the former, we simulate 4 independent channels, each with a capacity of 2 and a base failure rate of $0.001$. For the latter, we use a realistic LoRa simulator with a log-distance path loss.  A single base state located in the center receives data packets on 4 channels simultaneously. 

We consider that correcting the belief at each location has one unit of data value. Each action to send the data will consume one unit of energy for communication.  We compare the following methods:
\begin{itemize}
    \item \textbf{Random-p}: send the data when the belief is different from the real fire state with a fixed probability of $p\%$. 
    \item \textbf{Heuristic}: send the data whenever the belief at the current location is different from the real fire state.
    \item \textbf{IQL-S}: independent Q-learning with a softmax policy.
    \item \textbf{QMIX}~\cite{rashid2018qmix}: Q-learning based MARL algorithm for training decentralized policies with a centralized critic. 
    \item \textbf{DDPG}~\cite{lillicrap2019continuous}:  Deep Deterministic Policy Gradient to learn the optimal transmission probability. 
    \item \textbf{DDPG-FB}: DDPG with feedback on the channel utilization in the state. 
\end{itemize}

In addition, for the simple bandwidth-limited channel, a theoretical optimal distributed policy can be found as a performance upper bound:
\begin{itemize}
    \item \textbf{Optimal}: given the full information, the optimal stochastic policy obtained by solving \eqref{eq3}.
    \item \textbf{Optimal-Async}: the stochastic policy obtained by solving \eqref{eq3} based on past data values of other sensors, under the assumption that the  data values are not frequently made available to other sensors.
\end{itemize}

\begin{figure*}
    \centering
    \includegraphics[width=.95\linewidth]{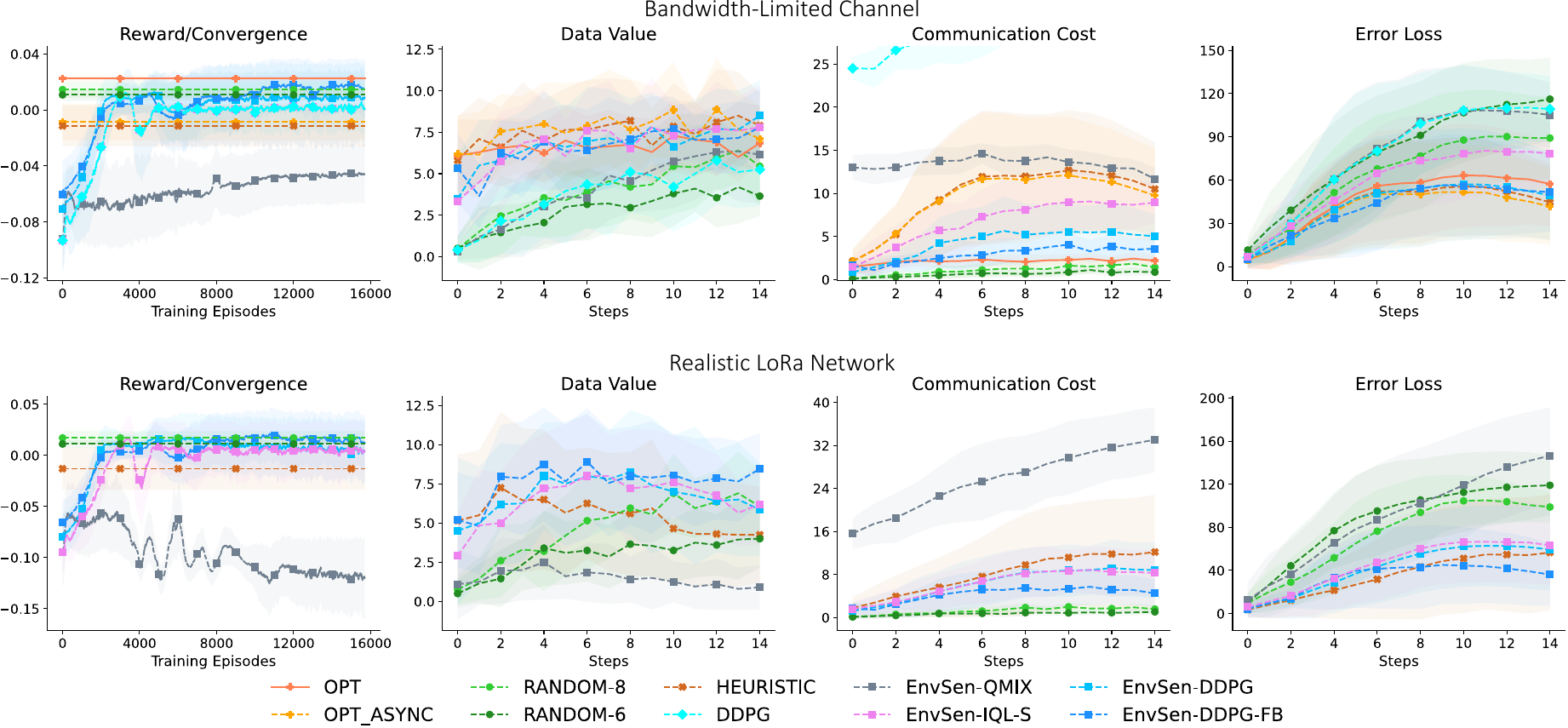}
    \caption{Experimental results. Columns from left to right: the average reward $v-wc$ with $w=0.2$/convergence of the policies, the sum of the data value, the total communication cost of all sensors, and the error loss of the belief for tracking wildfire, averaged across 30 random episodes. }
    \label{fig:experiments}
\end{figure*}
\begin{figure}
    \centering
 \subfloat[]
{\includegraphics[width=.5\linewidth]{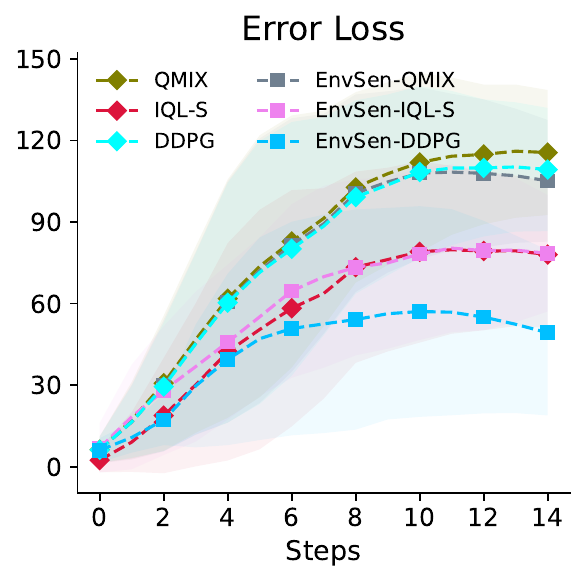}
\label{fig:data-value}}
 \subfloat[]
{\includegraphics[width=.5\linewidth]{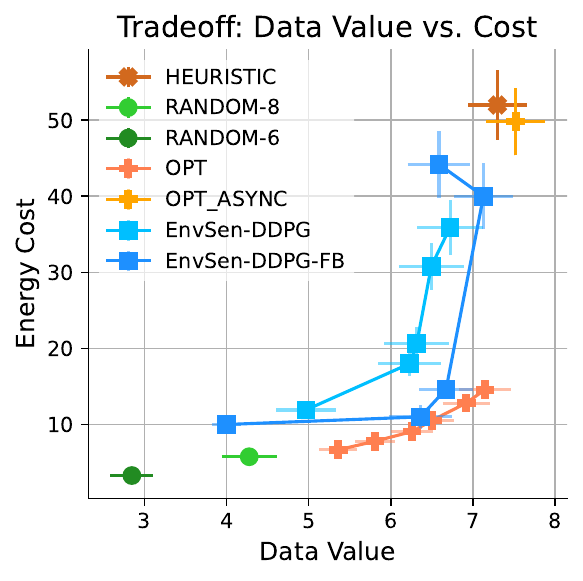}
\label{fig:tradeoff}}
\caption{(a): The error loss for tracking wildfire of the RL algorithms that directly use the tracking accuracy as the reward and those (EnvSen-X) that adopt our data metric. (b): the trade-off curve between data value and communication cost made by different policies. }
\label{fig:experiments2}
\end{figure}

\subsection{Evaluation}
We demonstrate the effectiveness of our defined data metric by comparing  QMIX, IQL-S, and DDPG with and without our environment sensing framework. The original algorithms for comparison directly use the tracking accuracy (i.e., the global objective) as the reward. Our EnvSen versions use the data value minus the communication cost (weighted by 0.2). The results, shown in Fig. \ref{fig:data-value}, indicate that the performance of DDPG is greatly improved with our framework, while both EnvSen-QMIX and EnvSen-IQL-S achieve slightly lower error losses. Our defined data value helps identify sensor data with higher value, leading to a more efficient reduction of error losses. Additionally, incorporating communication cost in the reward helps mitigate network congestion and enhances the throughput of valuable data.  

In Fig.~\ref{fig:experiments}, we evaluate different communication policies under  bandwidth-limited channels and realistic LoRa network conditions.  Our RL framework achieves low error loss with limited communication overhead, and approaches the performance of the optimal policy under bandwidth-limited channel conditions.  The heuristic policy maximizes data value but incurs high energy costs, while a random policy with low transmission probability reduces energy costs but leads to inefficient data collection and high error loss.  The performance of Optimal-Async is similar to the heuristic mainly because the data value of other sensors can be significantly underestimated with outdated information, leading to selfish behavior among sensors. This is especially true in fast-changing environments such as wildfires. EnvSen-QMIX consistently performs poorly and fails to converge possibly due to the complexity of training the centralized mixing network with a large number of agents. As a fully cooperative MARL method with a single global objective,  QMIX is unable to leverage individual feedback to assign credits effectively among agents.
 
Among the converging RL policies,  EnvSen-DDPG-FB achieves the highest return after convergence, maintaining high data value, low error loss, and moderate communication energy expenditure.  By leveraging feedback from the gateway, it improves upon EnvSen-DDPG's performance. Notably, despite obtaining similar rewards as the random policies, both EnvSen-DDPG-FB and EnvSen-DDPG achieve significantly lower error losses as they efficiently trade-off between data value and communication cost for a given weighting factor $w$. 

We investigate different weighting factors $w$ on the energy cost to analyze the trade-off between data value and cost. Fig. \ref{fig:tradeoff} shows the average energy cost and data value of each policy as we change $w$. The optimal policy achieves high data value while maintaining a low energy cost, maintaining a performance upper bound. The simple heuristic and random policies are insensitive to weight variations, while Optimal-Async performs similarly to the heuristic. EnvSen-DDPG adapts to different weights and approaches the optimal performance when provided with feedback.

\section{Discussion and Conclusion}\label{sec:conclusion}
In this work, we propose EnvSen, a MARL-based framework to find the optimal communication policies for a network of IoT sensors that track the states of an environment over time. 
Due to power and bandwidth limitations, each sensor must \emph{independently} decide whether to transmit its sensed data at each time step. However, the optimal sensing policy requires sensor coordination due to bandwidth constraints and spatiotemporal correlation in the data, even though the exact channel model and data correlation may need to be learned over time. Our EnvSen framework allows sensors to do so.

Our experiments on wildfire and LoRA wireless network simulators show that using EnvSen in RL algorithms performs favorably, as benchmarked by heuristics derived from the optimal communication policies in simplified settings. While we concentrate on the wildfire tracking problem in this work, we expect these communication policies can be applied to other environmental tracking problems, e.g., tracking air pollution's spread throughout a city, which also features spatiotemporal correlation in the sensed data. Our future work will also explore new definitions of the ``value'' of sensor data, e.g., based on known models of spatiotemporal diffusion for particular environmental tracking problems.

\section*{Acknowledgements}

This work was partially supported by the Office of Naval Research (N000142112128) and a Northrop Grumman gift.

\bibliographystyle{IEEEtran}
\bibliography{IEEEexample,reference}

\end{document}